\documentclass[a4paper]{article}

\usepackage{INTERSPEECH2022}
\usepackage[noend]{algpseudocode}
\usepackage{multirow}

\usepackage{algorithmicx,algorithm,cite}
\usepackage{threeparttable}
\usepackage{setspace}
\ninept

\title{A Comparative Study on Speaker-attributed Automatic Speech Recognition in Multi-party Meetings}
\name{
\begin{tabular}{c}
    \it Fan Yu$^1$, Zhihao Du, Shiliang Zhang, Yuxiao Lin$^2$, Lei Xie$^{1*}$\thanks{*Lei Xie is the corresponding author} \\
\end{tabular}
}

\address{
  $^1$Audio, Speech and Language Processing Group (ASLP@NPU), School of Computer Science,\\ Northwestern Polytechnical University, Xi’an, China \\
  $^2$College of Computer Science and Technology, Zhejiang University, Hangzhou, China}

\email{fyu@npu-aslp.org, lxie@nwpu.edu.cn}

\begin{document}

\maketitle
\begin{abstract}

In this paper, we conduct a comparative study on speaker-attributed automatic speech recognition (SA-ASR) in the multi-party meeting scenario, a topic with increasing attention in meeting rich transcription. Specifically, three approaches are evaluated in this study. 
The first approach, \textit{FD-SOT}, consists of a frame-level diarization model to identify speakers and a multi-talker ASR to recognize utterances. The speaker-attributed transcriptions are obtained by aligning the diarization results and the recognized hypotheses. 
However, due to the modular independence, such an alignment strategy may suffer from erroneous timestamps which severely hinder the model performance. 
Therefore, we propose the second approach, \textit{WD-SOT}, to address alignment errors by introducing a word-level diarization model, which can get rid of such timestamp alignment dependency.
To further mitigate the alignment issues, we propose the third approach, \textit{TS-ASR}, which trains a target-speaker separation module and an ASR module jointly.
By comparing various strategies for each SA-ASR approach, experimental results on a real meeting scenario corpus, AliMeeting, reveal that the WD-SOT approach achieves 10.7\% relative reduction on averaged speaker-dependent character error rate (SD-CER), compared with the FD-SOT approach.
In addition, the TS-ASR approach also outperforms the FD-SOT approach and brings 16.5\% relative average SD-CER reduction.

\end{abstract}

\vspace{-0.2cm}
\section{Introduction}
Speaker-attributed automatic speech recognition (SA-ASR) is the major purpose of rich transcription in real-world multi-party meetings~\cite{fiscus2006rich,fiscus2007rich}. In general, an SA-ASR aims at answering the question ``who spoke what''~\cite{BarkerWVT18,watanabe20b_chime,ryant2020third,mccowan2005ami}.
Compared with the multi-speaker ASR~\cite{yu2017recognizing,chen2017progressive,kanda2020serialized}, SA-ASR not only focuses on transcribing multi-speaker speech that may contain overlapped segments from different speakers, but also assigns speaker labels to each recognized word. 
Therefore, SA-ASR system needs to take more consideration of all involved modules, such as speaker diarization~\cite{park2022review,fujita2019end2,horiguchi2020end} to count and identify speakers, speech separation~\cite{yu2017permutation,hershey2016deep,chen2017deep} to handle overlapping speech and ASR~\cite{DBLP:journals/corr/VaswaniSPUJGKP17,gulati2020conformer,li2021recent} to recognize speech contents from the separated signals.

The accuracy of SA-ASR is affected by both transcript prediction and speaker assignment. Recently, a lot of efforts have been made on designing an end-to-end system that directly outputs multi-speaker transcriptions~\cite{yu2017recognizing,WatanabeRHSH18,ChenYLZMLWXL20,chang2019mimo,kanda2020serialized}.
Speech separation and joint-training with multiple ASR decoders under permutation invariant training (PIT) scheme is the typical approach~\cite{yu2017recognizing,WatanabeRHSH18,ChenYLZMLWXL20,chang2019mimo}. However, the maximum number of speakers that the model can handle is constrained by the number of the decoders in the model. Besides, duplicated hypotheses can be generated in different outputs, since the outputs are independent of each another in PIT. 
To mitigate these issues, the serialized output training (SOT) strategy~\cite{kanda2020serialized} is proposed for multi-talker ASR, which introduces a special symbol to represent the speaker change with only one output layer. In this way, SOT-based models have no constraints on the maximum number of speakers and avoid duplicated hypotheses naturally. In the recent M2MeT challenge which focuses on multi-speaker ASR in the meeting scenario~\cite{Yu2022M2MeT,Yu2022Summary}, SOT has been well applied and achieved remarkable performance. Therefore, in this paper, we combine the multi-speaker hypotheses of the SOT-based ASR model and the frame-level results of the speaker diarization model (e.g, TS-VAD~\cite{medennikov2020target,he2021target}) to obtain the speaker-attributed transcriptions by simply aligning the timestamps as our first approach, namely frame-level diarization with SOT (\textbf{FD-SOT}).

However, due to the modular independence, such an alignment strategy may suffer from erroneous timestamps which severely hinder the  speaker assignment performance. 
Considering that the lack of correlation between the ASR and speaker diarization modules may cause potential alignment errors, the recently updated revisions of SOT introduced speaker inventory and speaker encoder to produce speaker labels for each ASR token~\cite{KandaGWMCZY20,KandaYGWMCY21,kanda2021comparative}. Although they significantly reduced the speaker-attributed word error rate, these approaches cost a lot for changing ASR model structures and training pipelines, which is not practical in real production systems. 
In order to retain the original ASR network structure designed for single-speaker while addressing the alignment errors, we propose the second approach named word-level diarization with SOT (\textbf{WD-SOT}), which utilizes the recognized transcriptions from an SOT-based ASR to model the diarization system, to get rid of such timestamp alignment dependency. Meanwhile, we adopt self-attention to capture more contextual information, which can further improve the diarization performance.

Both FD-SOT and WD-SOT depend on the output of the SOT model, and the errors of SOT will seriously affect the performance of the overall framework.
Therefore, we turn to another solution of SA-ASR that uses a separation model to handle overlapped speech while getting rid of the dependence on the multi-talker ASR output~\cite{luo2019conv,wang2019voicefilter,yu2017recognizing,WatanabeRHSH18,chen2020continuous,wu2020end}. As the speech separation models are usually trained based on the signal-level criterion, a joint training strategy was proposed to mitigate the mismatch between the separation and the backend ASR, leading to better recognition performance~\cite{yu2017recognizing,WatanabeRHSH18,chen2020continuous,wu2020end}. However, most of the prior works adopted blind source separation (BSS), which cannot determine the specific speaker of the recognition transcriptions.
In order to fit the SA-ASR task, we use target speech separation (TSS) by adopting speaker diarization and speaker extraction models to obtain the speaker embeddings required by the TSS module. In this paper, we name this approach as joint target-speaker separation and ASR (\textbf{TS-ASR}) approach.

Although most of the works described above show promising results in multi-speaker ASR, they are mostly evaluated by simulated multi-speaker data. Problems including unknown number of speakers, variable overlap rate, and accurate speaker identity are still considered unsolved, especially for real-world applications. SA-ASR for multi-speaker speech remains to be a very difficult problem especially when speech utterances from multiple speakers significantly overlap in monaural recordings. 
To the best of our knowledge, we are the first to evaluate SA-ASR approaches on a real meeting corpus AliMeeting to provide reasonable SA-ASR results and promote the research in meeting rich transcription.

\vspace{-0.1cm}
\section{Evaluation Data}
\label{Evaluation}

In this study, we use the AliMeeting corpus~\cite{Yu2022M2MeT,Yu2022Summary} to evaluate various SA-ASR systems. Collected in real meetings, the AliMeeting corpus contains 104.75 hours~\footnote{Hours are calculated in a single channel of audio.} data for training (Train), 4 hours for evaluation (Eval) and 10 hours for test (Test). Each set contains several meeting sessions and each session consists of a 15 to 30-minute discussion with 2 to 4 participants. To highlight the speaker overlap, the sessions with 4 participants account for 59\%, 50\% and 57\% sessions in Train, Eval and Test, respectively. Train and Eval sets contain the 8-channel audios recorded from the microphone array (\textit{Ali-far}) as well as the near-field audio (\textit{Ali-near}) from the participant's headset microphone, while the Test set only contains the far-field audios. \textit{Ali-far-bf} is produced by applying CDDMA Beamformer~\cite{huang2020differential,zheng2021real}. In this paper, the model training and evaluation are all based on the single-channel audio, namely \textit{Ali-near} and \textit{Ali-far-bf}. The prefix \textit{Train-}, \textit{Eval-} and \textit{Test-} are used to indicate different sets, e.g., \textit{Train-Ali-far-bf} denotes the one channel data outputted by the beamformer which takes the AliMeeting 8-channel array Train data as input. We use official scripts~\footnote{https://github.com/yufan-aslp/AliMeeting} provided by the M2MeT challenge to prepare the sentence segmentation timestamp.
Meanwhile, in order to improve the performance of the speech separation module used in this paper, we simulate 50 hours mixed training data named \textit{Train-Ali-simu} from \textit{Train-Ali-near}.


\vspace{-0.1cm}
\section{SA-ASR}
\label{SA-ASR}

\vspace{-0.1cm}
\subsection{SOT}
\label{SOT}


The SOT method has an excellent ability to model the dependencies among outputs for different speakers and no longer has a limitation on the maximum number of speakers. To recognize multiple utterances, SOT serializes multiple references into a single token sequence with a special token $\langle \text{sc} \rangle$, which is used to concatenate the transcriptions of the different utterances. 
In order to avoid the complex calculation of PIT for all possible concatenation patterns, SOT sorts the reference labels by their start times, which means “first-in, first-out” (FIFO). The experiments show that the FIFO method achieves a better CER than the method of calculating all permutations \cite{kanda2020serialized}. Meanwhile, considering the high overlap ratio and frequent speaker turn on the AliMeeting corpus, we experimentally investigate the speaker-based and utterance-based FIFO training schemes~\cite{kanda2021investigation} in section \ref{Experiment}. 
Remarkably, we employ Transformer~\cite{DBLP:journals/corr/VaswaniSPUJGKP17} model with Conformer encoder~\cite{gulati2020conformer}, 
which includes 12 encoder layers and 6 decoder layers. The common parameters of the encoder and decoder layers are: $d^{head} = 4, d^{attn} = 256, d^{ff} = 2048$ for the number of attention heads, dimension of attention module, and dimension of feedforward layer, respectively.



\begin{figure}[t]
 	\centering
 	\includegraphics[width=1.0\linewidth]{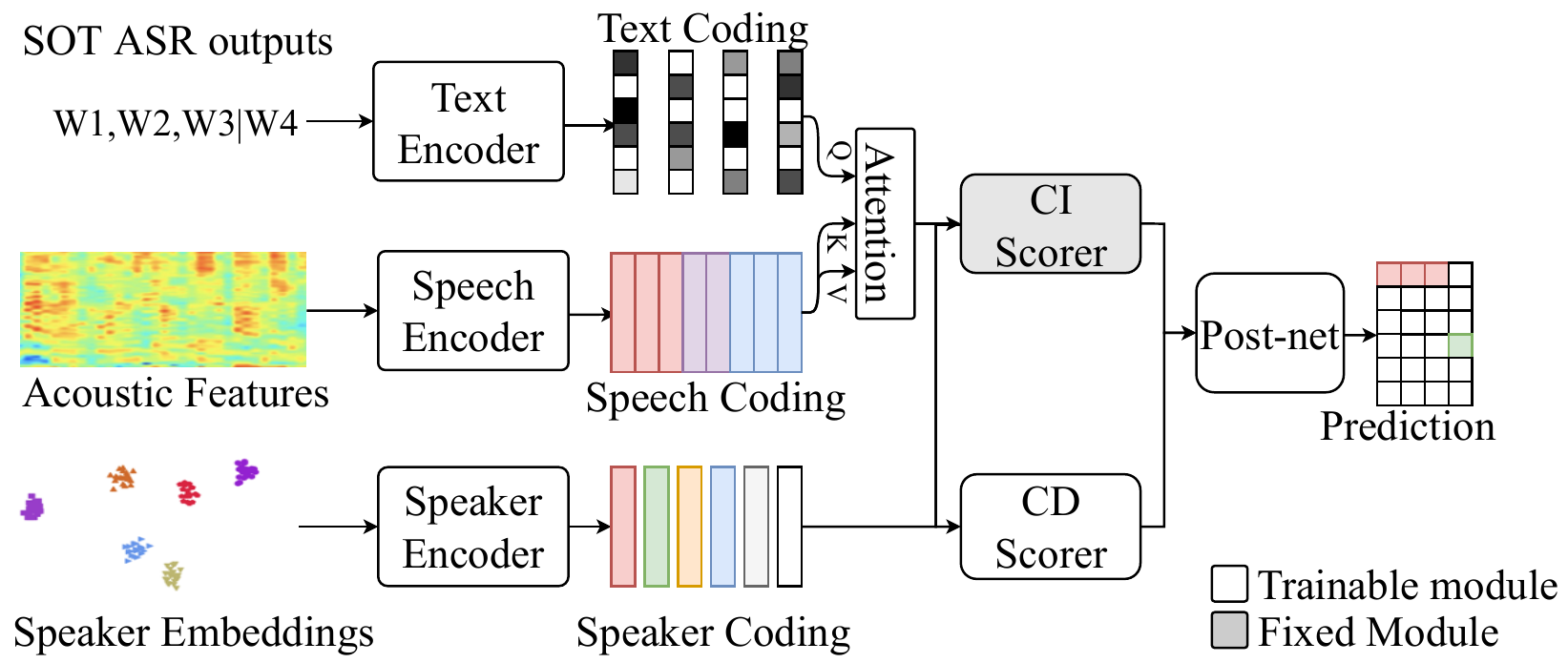}
 	 \vspace{-0.5cm}
 	\caption{
 	 Semantic diagram of the proposed world-level diarization method.
 	}
 	\label{sys_systure_send}
 	\vspace{-0.7cm}
\end{figure}

\vspace{-0.1cm}
\subsection{FD-SOT}
\label{FD-SOT}

As the top three teams all employ TS-VAD to find the overlap between speakers in the M2MeT challenge speaker diarization track~\cite{Yu2022Summary}, we also re-implemented it and achieved DERs of 4.20\% and 5.42\% on AliMeeting Eval and Test sets, respectively.
To further obtain the speaker-attributed transcriptions, we combine the results of TS-VAD and SOT by aligning the timestamps. This approach is named as frame-level diarization with SOT (FD-SOT).
The detailed process of FD-SOT is showing as follows:

\vspace{-0.1cm}
\begin{itemize}
\setlength{\itemsep}{0pt}
\setlength{\parsep}{0pt}
\setlength{\parskip}{0pt}
	\item[1)] We estimate the number of the utterances using the oracle sentence segmentation, says $\hat{N}$, according to the diarization output of TS-VAD. 
	\item[2)] The utterance number of SOT output is defined as $N$. If $\hat{N}$ is equal to $N$, no further effort is required.
	\item[3)]  If $\hat{N}$ is larger than $N$,  we select $N$ out of the utterances that have the longest duration in the TS-VAD diarization output, and discard other utterances. 
	\item[4)]  If $\hat{N}$ is smaller than $N$, we select $\hat{N}$ out of utterances that have the longest text length in the SOT output, and discard other utterances. 
	\item[5)] Finally, we match utterances between TS-VAD and SOT in chronological order.
\end{itemize}


\begin{figure*}[t]
	\centering
	\vspace{-7pt}

	\includegraphics[width=0.90\linewidth]{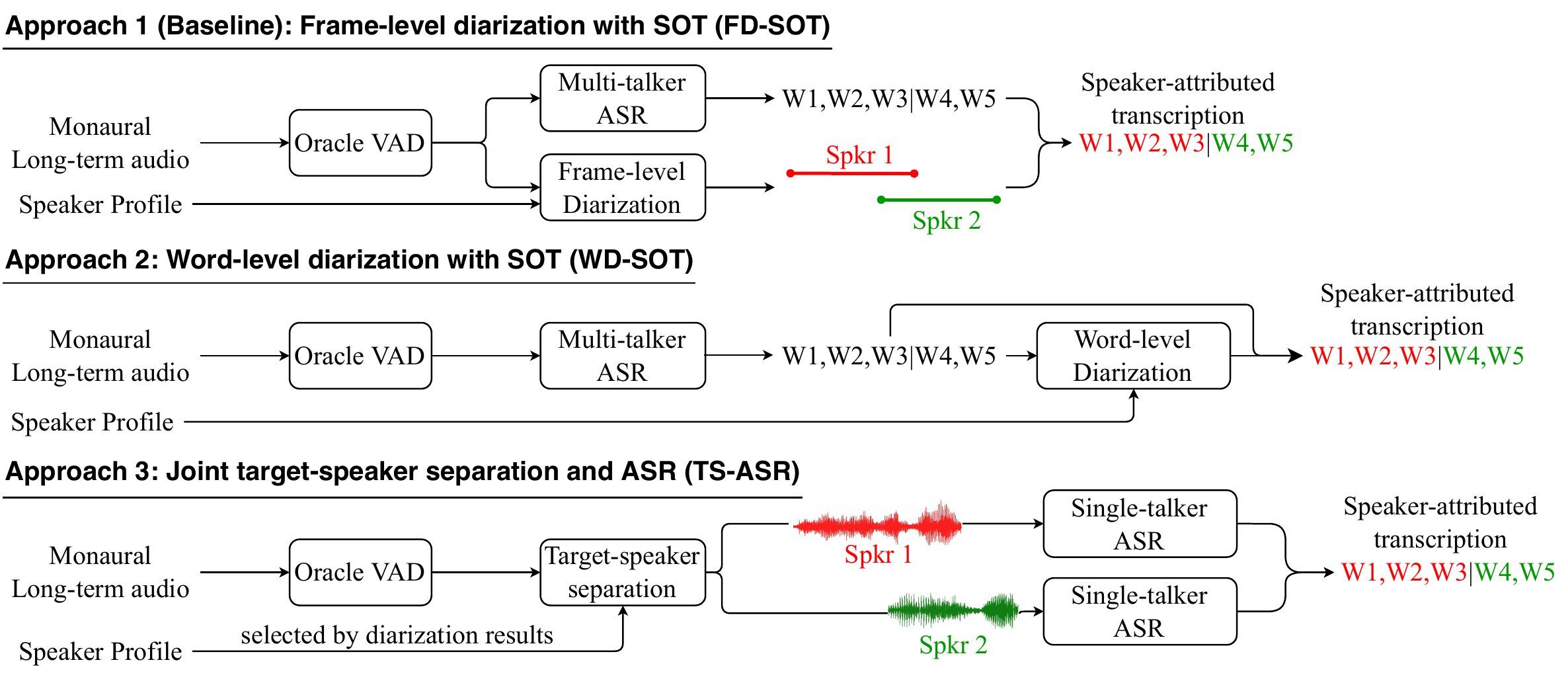}
	\vspace{-0.5cm}
	\caption{
	 Semantic diagram of compared approaches, including FD-SOT, WD-SOT and TS-ASR.
	}
	\vspace{-0.3cm}
	\label{sys_systure}
	\vspace{-0.4cm}
\end{figure*}

\vspace{-0.4cm}
\subsection{WD-SOT}

In the proposed world-level diarization (WD) method, as shown in Fig.~\ref{sys_systure_send}, we first use three individual encoders to encode the multi-talker hypotheses, speech features and speaker embeddings. Given the encoded hypotheses $H=\{\mathrm{h}_l|l=1,\dots,L\}$ and features $X=\{\mathrm{x}_t|t=1,\dots,T\}$, a multi-head attention is used to produce the aggregated feature representation $\mathrm{r}_l$ for each token:
\begin{equation}
	\mathrm{r}_l = \sum_{t=1}^{T}{a_{l,t}W_v\mathrm{x}_t}
\end{equation}
which is thought to include both acoustic and semantic information. $a_{l,t}$ is calculated by 
\begin{equation}
	\begin{split}
		\alpha_{l,t} &= \text{dot}(W_q \mathrm{h}_l, W_k \mathrm{x}_t), \\
		a_{l,t} &= \frac{\exp(\alpha_{l,t})}{\sum_{i=1}^{T}\exp(\alpha_{l,i})}.
	\end{split}
\end{equation}
$W_q$, $W_k$ and $W_v$ are trainable parameters in the attention layer. Next, the context-independent (CI) score $S^{CI}$ is derived from the dot product between the encoded speaker embeddings $\mathbf{v}_n$ and the aggregated representations $\mathbf{r}_l$:

\begin{equation}
	S^{CI}_{l,n} = \text{dot}(\mathrm{r}_l, \mathrm{v}_n).
\end{equation}
While CI scores only consider the representations of the current speaker, the contextual information of different speakers is also useful to identify the activated speaker from others.
Therefore, we further design a context-dependent (CD) score $S^{CD}_{l,n}$, which is defined as follows:

\vspace{-0.2cm}
\begin{equation}
	S^{CD}_{l,n} = f(\mathbf{r}_l, \mathbf{v}_n; R,\Theta)
\end{equation}
where $f$ is a context-aware function, e.g., the self-attention based networks (SAN)~\cite{DBLP:journals/corr/VaswaniSPUJGKP17}, and $\Theta$ are learnable parameters of $f$. $R=\{\mathbf{r}_l|l=1,\dots,L\}$ contains all aggregated representations in an utterance.
Finally, the CI and CD scores are concatenated and fed to a post-processing network to predict the corresponding speaker for each character.

In this study, a four-layer self-attention based encoder is employed to encode the recognized text with 8 attention heads and 256 hidden units in each layer. 

\vspace{-0.2cm}
\subsection{TS-ASR}
\vspace{-0.1cm}

Target-speaker separation modules generate target-speaker representation from multi-speaker signals by enrollment embedding for ASR.  With the premise that optimizing front-end and back-end separately will lead to sub-optimal performance, joint modeling will make the whole system matching the final metric. 
Meanwhile, we adopt TS-VAD described in Section~\ref{FD-SOT} and d-vector extraction network to obtain the speaker embeddings required by the target-speaker separation module.

For the structure of target-speaker separation model , we compared the performance between Conformer~\cite{gulati2020conformer} and convolutional recurrent network (CRN)~\cite{tan2018convolutional}. Target-speaker separation Conformer network consists of 6 encoder layers with 4 attention heads, 256 attention dimensions and 2048 dimensional feed-forward network. CRN network consists of 2 bi-directional long short-term memory (BLSTM) layers with 256 hidden dimensions and a convolutional encoder-decoder (CED) with skip connections of corresponding encoder and decoder layer.
We employ a Res2Net-based d-vector extraction network trained on the VoxCeleb corpus~\cite{nagrani2017voxceleb,chung2018voxceleb2} as the speaker embedding model. Noted that the d-vector is combined with both front-end models using feature-wise linear modulation (FILM)~\cite{perez2018film}, which shows better performance on the multi-speaker and noisy datasets. 
In order to control the TS-ASR model parameters to be consistent with the ASR model of previous approaches, we adopt the ASR model with 6 layers of encoder.


\vspace{-0.2cm}
\section{Experiments}
\label{Experiment}

\vspace{-0.1cm}

\subsection{Training details}
\vspace{-0.1cm}

In our work, the 80-dimensional log Mel-filter bank feature (Fbank) is used as the input feature. The window size is 25 ms with a shift of 10 ms. We use 4950 Chinese characters extracted from the training transcriptions as the modeling units. 
SOT models are trained directly based on \textit{Train-Ali-far-bf}, \textit{Train-Ali-near} and \textit{Train-Ali-simu}. For front-end and back-end joint models, we first pre-train each module with \textit{Train-Ali-simu} and \textit{Train-Ali-near}, respectively. Then we use the \textit{Train-Ali-far-bf} data for fine-tuning and use ASR loss function to update the parameters of the whole model for joint training. We train the model for 100 epochs with the Adam optimizer. 

\vspace{-0.2cm}
\subsection{Evaluation metric}
\vspace{-0.1cm}

We use two evaluation metrics in our experiments, referring as speaker independent- (SI-) and speaker dependent- (SD-) character error rate (CER)~\cite{fu2021aishell}. The SI-CER is designed to measure the performance of multi-speaker ASR task (e.g, track 2 of M2Met challenge~\cite{Yu2022M2MeT,Yu2022Summary}), which ignores the speaker labels. On the other hand, the SD-CER, slightly different from cpWER~\cite{kanda2021comparative}, is calculated by comparing the ASR hypothesis and the reference transcription of the corresponding speaker instead of all possible speaker permutations. The SD-CER is more rigorous in calculation due to the global calculation of the entire meeting, which needs to determine the exact global speaker ID for each utterance in the whole meeting.

\vspace{-0.2cm}
\subsection{Comparison of different SA-ASR approaches}
\vspace{-0.1cm}

As shown in Table~\ref{tab:result_pipline}, we evaluate our SA-ASR approaches on AliMeeting Eval and Test sets. We first compare the speaker-based and utterance-based FIFO training schemes of SOT models described in Section~\ref{SOT}. Consistent with the conclusion of~\cite{kanda2021investigation}, we find that the utterance-based FIFO training significantly outperforms the speaker-based FIFO training with over 1\% absolute SI/SD-CER reduction for all SOT-based approaches, due to the high overlap ratio and frequent speaker turns of AliMeeting. Based on this conclusion, we employ the utterance-based FIFO scheme in the remaining experiments.

For the SOT-based model, we regard the SI-CER result of SOT as the topline of SD-CER results, assuming that each token matches the correct speaker. We can see that our proposed WD-SOT approach outperforms the FD-SOT approach, leading to 12.2\% (41.0\% $\to$ 36.0\%) and 9.6\% (41.2\% $\to$ 37.1\%) relative SD-CER reduction on Eval and Test set, respectively. Compared with SOT-based SA-ASR models, our proposed TS-ASR models achieve the lowest averaged SD-CER. Specifically, TS-ASR (CRN) approach achieves SD-CERs of 32.5\% and 35.1\% on Eval and Test sets, respectively.


\vspace{-0.2cm}
\begin{table}[!htb]
\centering
\caption{SA-ASR results for various modular approaches on Eval and Test sets (\%).}
\vspace{-0.2cm}
\setlength{\tabcolsep}{5pt}
\begin{tabular}{c|c|ccc}
\toprule
\hline
Approach & Metric & Eval                      & Test           & Average           \\ \hline
SOT~\cite{Yu2022M2MeT,Yu2022Summary}                   &SI-CER         &  29.7      &  30.9    &   30.6        \\ \hline
FD-SOT                &\multirow{4}{*}{SD-CER}         &  41.0      &  41.2    &   41.2        \\
WD-SOT                &         &  36.0      &  37.1    &   36.8 \\
TS-ASR (Confomer)     &         & 34.8       &  \textbf{34.7}    &   34.7  \\
TS-ASR (CRN)  &         & \textbf{32.5}       &  35.1    &   \textbf{34.4}  \\
\hline
\bottomrule

\end{tabular}
\label{tab:result_pipline}
\vspace{-0.3cm}

\end{table}

\vspace{-0.3cm}
\subsection{Comparison of various strategies for WD-SOT}

We further compare the effect of various strategies for the WD-SOT approach, and the results are shown in Table~\ref{tab:result_wdsot}. The first-row result of WD-SOT is trained using the ground truth transcripts of \textit{Train-Ali-far-bf} only. To increase the robustness of the model, we add the hypothetical transcriptions to the training set, which dramatically decreases the average SD-CER from 39.1\% to 37.9\%.
From the comparison of contextual information, we can conclude that when the WD-SOT captures more context information through the self-attention layer, the performance can be significantly improved, which achieves 2.9\% (37.9\% $\to$ 36.8\%) relative average SD-CER reduction.
Our WD-SOT approach is based on the results of SOT, so the SOT performance will seriously affect the performance of the whole framework. 
Considering the position offset of the separator $\langle \text{sc} \rangle$, we intend to investigate the impact of the separator prediction accuracy on the model performance. After replacing the oracle separator, the overall performance has been improved from 36.8\% to 36.3\%, especially on Test sets with 2.5\% relative SD-CER reduction.

\vspace{-0.2cm}
\begin{table}[!htb]
\centering
\caption{Comparison of the WD-SOT results of various strategies on Eval and Test sets (SD-CER \%).}
\vspace{-0.2cm}
\begin{tabular}{lccc}
\toprule
\hline
Approach      & Eval             & Test      & Average          \\ \hline
WD-SOT                             & 40.6     & 38.5    &    39.1                \\
\quad + Hypothetical transcriptions                               & { 37.8} & { 38.0}    & 37.9 \\
\quad  \quad  + Contextual  information                                 & \textbf{ 36.0} & { 37.1} & 36.8 \\
\quad  \quad  \quad + Oracle separator $\langle \text{sc} \rangle$ &    {36.4}  &	 \textbf{36.2}           &  \textbf{36.3}    \\
\hline
\bottomrule
\end{tabular}
\label{tab:result_wdsot}
\vspace{-0.3cm}

\end{table}

\vspace{-0.3cm}
\subsection{Impact of minimum time of diarization utterances for TS-ASR approach}

In the TS-ASR approach, we need to determine the speakers within an oracle sentence segment, which relies on two estimated information: the oracle speaker labels and the speaker diarization results. Surprisingly, we obtain better recognition performance by using speaker diarization results (0s) for both TS-ASR approaches, comparing with the oracle speaker labels, leading to 7.0\%/8.2\% and 2.5\%/3.4\% relative SD-CER reduction on Eval and Test set for Conformer and CRN TS-ASR approaches, as shown in Table~\ref{tab:result_tsasr_eval}. By analyzing the decoding results, we find that some interfering speech is recognized when target-speaker speech duration is short, resulting in a large number of insertion errors.
Compared with the insertion errors caused by oracle speaker labels covering all speaker speech,  the deletion errors caused by  speaker diarization results  ignoring short speaker speech are fewer.
Based on this finding, we further investigate the impact of minimum time of diarization utterances for TS-ASR approaches. From the table, we can see that both TS-ASR models achieve the best results at minimum time equal to 0.5 s, which brings absolute SD-CER reduction ranging from 0.5\% to 0.8\% on the Eval and Test sets compared with the TS-ASR approaches without deleting short speaker speech.


\vspace{-0.2cm}
\begin{table}[!htb]
\setlength{\tabcolsep}{5pt}
\centering
\caption{TS-ASR results of the different minimum time of diarization utterances on Eval and Test sets (SD-CER \%).}
\vspace{-0.2cm}

\begin{tabular}{ccccccc}
\toprule
\hline
\multicolumn{1}{c}{}                         &           &                  & \multicolumn{4}{c}{Speaker diarization}                                                                               \\ \cline{4-7} 
\multicolumn{1}{c}{\multirow{-2}{*}{Approach}}  & \multirow{-2}{*}{Set} & \multirow{-2}{*}{Oracle}    & 0s                          & 0.3s                        & 0.5s                        & 0.7s                        \\ \hline
Confomer         & \multirow{2}{*}{Eval}                    & { 37.4} & { 34.8} & { 34.4} & { \textbf{34.3}} & { 34.3} \\
CRN      &                    & { 35.4} & { 32.5} & { 32.0} & { \textbf{31.9}} & { 31.9} \\ \hline
Confomer         & \multirow{2}{*}{Test}                    & { 35.6} & { 34.7} & { 34.3} & { \textbf{34.2}} & { 34.3} \\
CRN      &                    & { 36.3} & { 35.1} & { 34.5} & { \textbf{34.3}} & { 34.4} \\
\hline
\bottomrule

\end{tabular}
\label{tab:result_tsasr_eval}
\vspace{-0.3cm}
\end{table}


\vspace{-0.3cm}
\subsection{Effect of joint training for TS-ASR approach}

The comparison results between Conformer and CRN TS-ASR approaches with different optimization strategies are shown in Table~\ref{tab:result_tsasr_comp}. Here, the front-end modules are pre-trained on \textit{Train-Ali-simu} set and back-end modules are pre-trained with \textit{Train-Ali-near} data, respectively. The difference between the separated and joint training strategies is whether we use ASR loss function to update the front-end module when we use \textit{Train-Ali-far-bf} data to fine-tune the whole model.
According to the Table~\ref{tab:result_tsasr_comp}, joint optimization for Conformer and CRN TS-ASR approaches leads to 26.8\% (47.4\% $\to$ 34.7\%) and 23.9\% (45.1\% $\to$ 34.3\%) relative average SD-CER reduction on Eval and Test set, respectively. We conclude that the joint optimization can make the front-end module more suitable and less distorted for the back-end ASR.

\vspace{-0.2cm}
\begin{table}[!htb]
\setlength{\tabcolsep}{5pt}
\centering
\caption{The comparison of separated v.s. joint optimization for TS-ASR on Eval and Test sets (SD-CER \%).}
\vspace{-0.2cm}
\begin{tabular}{ccccc}
\toprule
\hline
Approach & Optimize strategy & Eval             & Test &Average             \\ \hline
\multirow{2}{*}{Conformer}                                  & Separated        & 46.0                        & 48.0  &  47.4                    \\
                                  & Joint           & { 34.8} & \textbf{34.7}    &  34.7 \\ \hline
\multirow{2}{*}{CRN}                                & Separated        & 43.3                        & 45.8          & 45.1              \\
                               & Joint           & \textbf{32.5} & { 35.1} & \textbf{34.3} \\ 
\hline
\bottomrule
\end{tabular}
\label{tab:result_tsasr_comp}
\vspace{-0.2cm}

\end{table}

\vspace{-0.5cm}
\section{Conclusion}
\label{Conclusions}
\vspace{-0.2cm}

In this study, three SA-ASR approaches are evaluated on the AliMeeting corpus, a challenging meeting dataset with multi-talker conversation. Compared with the baseline approach, FD-SOT, the proposed WD-SOT approach addresses the alignment errors by introducing a word-level diarization model and results in 10.7\% relative average SD-CER reduction. 
To further get rid of the dependence on multi-talker ASR output, the proposed TS-ASR approach trains a target-speaker separation module and an ASR module jointly, which leads to 16.5\% relative average SD-CER reduction compared with FD-SOT.
Moreover, ignoring short diarization utterances can bring 0.8\% absolute SD-CER reduction for the TS-ASR task.
In the future, we will investigate how to incorporate the single-speaker ASR trained on large-scale data into our proposed approaches for real-world applications.
\vspace{-0.3cm}

\section{Acknowledgement}
\vspace{-0.2cm}

This work was supported by Alibaba Group through Alibaba Research Intern Program and Key R \& D Projects of the Ministry of Science and Technology (2020YFC0832500).

\begin{spacing}{0.01}
\ninept
\bibliographystyle{IEEEtran}
\bibliography{bib_new}
\end{spacing}

\end{document}